\documentclass[conference]{IEEEtran}
\IEEEoverridecommandlockouts
\usepackage{cite}
\usepackage{amsmath,amssymb,amsfonts}
\usepackage{algorithmic}
\usepackage{graphicx}
\usepackage{textcomp}
\usepackage{xcolor}
\usepackage{url}
\def\BibTeX{{\rm B\kern-.05em{\sc i\kern-.025em b}\kern-.08em
    T\kern-.1667em\lower.7ex\hbox{E}\kern-.125emX}}

\usepackage{graphicx}
\usepackage{stfloats}
\usepackage{graphicx}
\usepackage{tikz}
\usetikzlibrary{arrows.meta, positioning}
\usepackage{eso-pic}
\usepackage{graphicx}

\ifCLASSOPTIONcompsoc
    \usepackage[caption=false, font=normalsize, labelfont=sf, textfont=sf]{subfig}
\else
    \usepackage[caption=false, font=footnotesize]{subfig}
\fi

\begin{document}

\title{Parameters Overshadowed by Price Lags: Load, Climate, and Calendar Effects in ERCOT Day-Ahead Price Formation}
% \thanks{add TUBITAK and iSTAR}
% }
% "Beyond Price Lags: Quantifying Variable Masking in ERCOT Day-Ahead %Markets via Suppression Ratio Analysis"
% "Isolation of Physical Drivers in Electricity Price Formation: A %Suppression Ratio Approach using HGBR"
% "The Masking Effect of Lagged Prices: Measuring Feature Suppression in %Day-Ahead Market Forecasting"
\author{
\IEEEauthorblockN{Yunus Bicen, Eman Hammad}
\IEEEauthorblockA{
iSTAR Lab, Texas A\&M University, College Station, TX, USA\\
\{yunusbicen, eman.hammad\}@tamu.edu
}
}

% \author{\IEEEauthorblockN{1\textsuperscript{st} Yunus Bicen}
% \IEEEauthorblockA{\textit{Engineering Technology \& Industrial Distribution} \\
% \textit{Texas A\&M University}\\
% College Station, TX, USA \\
% yunusbicen@tamu.edu}
% \and
% \IEEEauthorblockN{2\textsuperscript{nd} Eman Hammad}
% \IEEEauthorblockA{\textit{Engineering Technology \& Industrial Distribution} \\
% \textit{Texas A\&M University}\\
% College Station, TX, USA \\
% eman.hammad@tamu.edu}

% }
\AddToShipoutPictureFG*{%
  \AtPageLowerLeft{%
    \raisebox{5mm}{%
      \hspace{8mm}%
      \parbox{0.96\paperwidth}{%
        \fontsize{5}{6}\selectfont
        \textcopyright~2026 IEEE. Personal use of this material is permitted.
        Permission from IEEE must be obtained for all other uses, in any current
        or future media, including reprinting/republishing this material for
        advertising or promotional purposes, creating new collective works,
        for resale or redistribution to servers or lists, or reuse of any
        copyrighted component of this work in other works.
        
        \vspace{1mm}
        \textit{Accepted author manuscript at the International Conference on Communications, Control, and Computing Technologies for Smart Grids 2026 (IEEE SmartGridComm 2026). This version has been
        accepted for publication and may differ slightly from the final
        published version.}
      }%
    }%
  }%
}

\maketitle

\begin{abstract}
%This study investigates the underlying physical and structural mechanisms masked by lagged price variables, as well as how various factors influence market structure. To this end, the ERCOT Day-Ahead Market was analyzed by distinguishing between normal and price-spike regimes, and by evaluating the contributions of load, climatic, and temporal variables both with and without the inclusion of lagged prices. It was observed that, particularly during periods of price spikes, neither price lags nor load demand were sufficient to explain the observed market dynamics. Conversely, on the physical side, nonlinear interactions between temperature and humidity were found to form a distinct pattern. Furthermore, the study revealed that price spikes are not merely the result of continuously rising temperatures, but rather emerge from the system reacting sharply within a specific temperature range. These findings underscore the importance of moving beyond lag-based models to uncover the true drivers of extreme price events.

Accurate electricity price forecasting is critical for smart grid stability, yet the heavy reliance on historical price lags in modern predictive models often obscures the fundamental physical drivers of market volatility. This paper proposes a regime-sensitive, explainable artificial intelligence (XAI) framework to unmask the hidden roles of load, climate, and calendar variables in the ERCOT Day-Ahead Market (2014–2024). Utilizing a Histogram-based Gradient Boosting Regressor (HGBR), we introduce a Suppression Ratio to quantify how price lags overshadow physical parameters. Our analysis reveals that while lags provide short-term memory during normal conditions, they fail to capture extreme dynamics; notably, their inclusion increased forecasting error (MAE/RMSE) during price spikes. By isolating these effects, the study revealed that price spikes are not merely the result of continuously rising temperatures, but rather emerge from the system reacting sharply within a specific temperature range.
These findings underscore the importance of moving beyond lag-based models to uncover the true drivers of extreme price events.
\end{abstract}


\section{Introduction}

% The transition toward decarbonized smart grids has introduced unprecedented volatility into electricity markets. In high-renewable penetration zones the traditional relationship between load and price has been disrupted by variable generation and extreme weather events. Consequently, accurate Day-Ahead Market (DAM) price forecasting has become essential for grid stability, demand-side management, and the economic viability of energy storage systems.

Today, the operation of electrical energy within a competitive framework under free-market conditions, much like its physical management, has undergone significant changes in parallel with technological advancements \cite{b1,b2,b3}. Under competitive conditions, accurately forecasting market prices is of critical importance, along with efficient utilization of electrical energy \cite{b4,b5}. In an electricity market characterized by a large number of participants and a wide variety of variables, profit margins can be increased by appropriately structuring the relevant pricing strategy based on the physical conditions of future time periods \cite{b5,b6,b7}. For this reason, there has been a lot of focus on developing models with high predictive accuracy. In this regard, both the forecasting models examined in the literature and the input variables employed within them have become increasingly diverse \cite{b8}.

Traditional forecasting frameworks originally relied on stochastic models to identify local behavioral patterns in day-ahead market prices \cite{b8,b9}, with SARIMA and SARIMAX emerging as the most prevalent benchmarks \cite{b10,b11}. To address specific market complexities, researchers introduced Dynamic Harmonic Regression to capture multi-level seasonality \cite{b12}, Markov Regime-Switching models to handle price spikes \cite{b13}, and GARCH models for volatility modeling \cite{b14}. As the field evolved, artificial neural networks \cite{b15} and machine learning techniques gained prominence \cite{b16}, expanding the diversity of input features. Within this paradigm, Random Forest (RF) \cite{b17} and Gradient Boosting variants like XGBoost have become industry standards \cite{b18,b19}. The advent of deep learning further transformed forecasting, with Long Short-Term Memory (LSTM) networks becoming the dominant architecture due to their ability to capture long-term temporal dependencies \cite{b20}, often complemented by hybrid configurations such as LSTM-CNN \cite{b21}.

Despite these advances, forecasting error remains an inherent characteristic of electricity markets, shifting recent research focus toward quantifying uncertainty through probabilistic forecasting \cite{b22}. Quantile Regression, implemented via linear models, RF, or Gradient Boosting Machine(GBM), has become a cornerstone for generating prediction intervals essential for risk management \cite{b22,b23}, alongside emerging frameworks like Conformal Prediction \cite{b24}. Beyond mere predictive accuracy, there is a growing need to interpret the underlying drivers of specific model outputs. Consequently, Explainable AI (XAI) techniques, particularly SHAP-based importance and Permutation Feature Importance (PFI) analyses, have recently begun to redefine the transparency of electricity price formation \cite{b25,b26}.

%A general overview of the literature reveals that the primary objective of most studies is to enhance prediction accuracy. Studies aimed at truly understanding the price mechanism are limited, and methods such as SHAP are typically employed to interpret the model itself. In current approaches, lags, particularly price lags, tend to attenuate the intrinsic impact (explanatory power) of most parameters. Yet, the natural and fundamental determinants of market price are load, climatic parameters, and calendar tags. Consequently, in this study, we focused on investigating the isolated effects of these fundamental parameters on market behavior. The study was conducted using the Histogram-based Gradient Boosting Regressor (HGBR), a tree-based ensemble method that is highly suitable for the high-volume and nonlinear nature of hourly ERCOT day-ahead price data. We quantified the extent to which the influence of these natural parameters diminishes upon the inclusion of lags by calculating the Suppression Ratio (SR). These analyses were conducted under two distinct conditions: normal and spike market regimes. Thus, it was possible to observe the degree to which each specific parameter diverges within each respective regime. SR values were computed based on Permutation Feature Importance PFI rather than SHAP values, in order to capture the change in model reliance on each feature when price lag variables are introduced. This approach contributes to constructing a nonparametric, regime-sensitive, and interpretable framework.

A survey of the existing literature reveals that most recent research is predominantly driven by the objective of maximizing predictive accuracy, often at the expense of model interpretability. While techniques such as SHAP are increasingly employed to explain model behavior, few studies aim to fundamentally deconstruct the underlying price formation mechanism. A critical oversight in current methodologies is the over-reliance on historical price lags, which tend to attenuate the intrinsic explanatory power of physical drivers. In reality, the fundamental determinants of market prices are grounded in load demand, climatic conditions, and temporal calendar effects. This study addresses this gap by isolating the effects of these fundamental parameters to uncover their true impact on market dynamics. Using a Histogram-based Gradient Boosting Regressor (HGBR), an ensemble method uniquely suited for the high-volume, nonlinear nature of ERCOT day-ahead data, we introduce the Suppression Ratio (SR). This metric quantifies the extent to which physical parameters are overshadowed when price lags are introduced into the feature set. By bifurcating our analysis into "normal" and "spike" market regimes, we demonstrate how the reliance on specific drivers diverges under extreme conditions. Unlike traditional SHAP-based interpretations, our SR values are derived from Permutation Feature Importance (PFI) to precisely capture the shift in model dependency. This approach establishes a nonparametric, regime-sensitive framework that prioritizes the structural transparency of electricity markets.

% \begin{figure*}[!t]
% \centering
%     \centering
%     \includegraphics[width=0.65\textwidth]{fig01}
%     \vspace{-0.25cm} 
%     \caption{Daily maximum DA prices with 90-day rolling mean}

% %\vspace{-0.15cm}   % İki şekil arasında boşluk

%     \centering
%     \includegraphics[width=0.65\textwidth]{fig02}
%     \vspace{-0.25cm} 
%     \caption{Yearly price distributions (2014-2024)}
%     \label{fig:XXXb}

% \end{figure*}
\begin{figure*}[!t]
    \centering
    \begin{minipage}{0.48\textwidth}
        \centering
        \includegraphics[width=\linewidth]{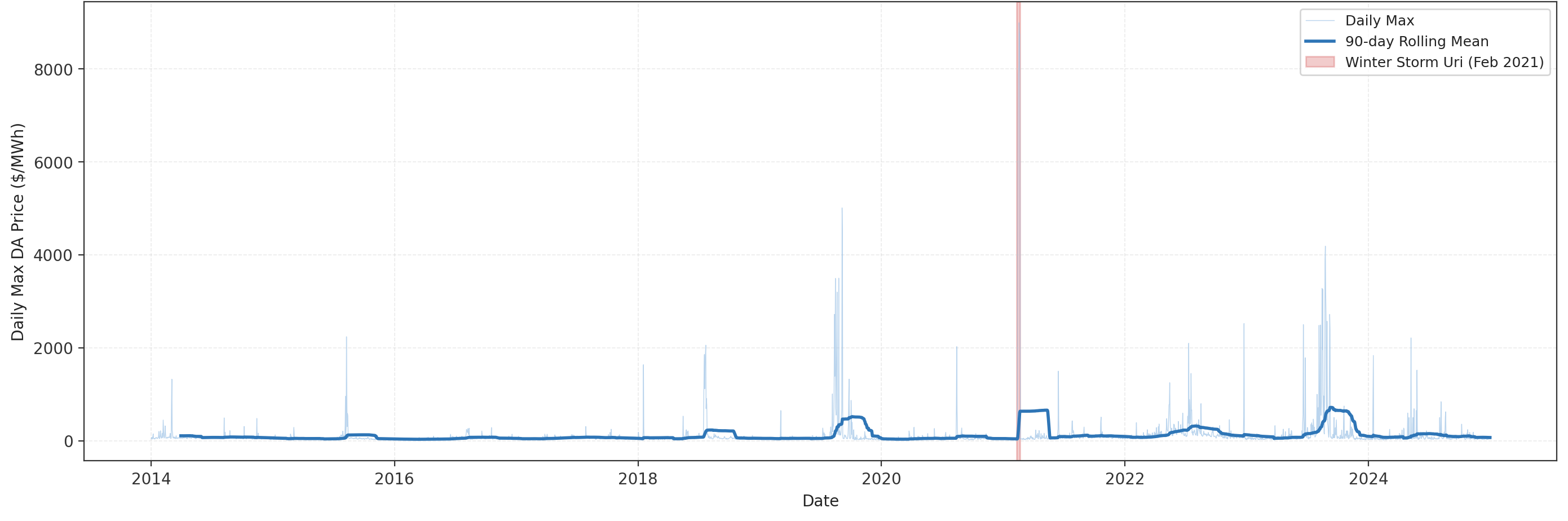}
        \vspace{-0.2cm}
        \caption{Daily maximum DA prices with 90-day rolling mean}
        \label{fig:daily_max}
    \end{minipage}
    \hfill % Adds horizontal spacing between the two figures
    \begin{minipage}{0.48\textwidth}
        \centering
        \includegraphics[width=\linewidth]{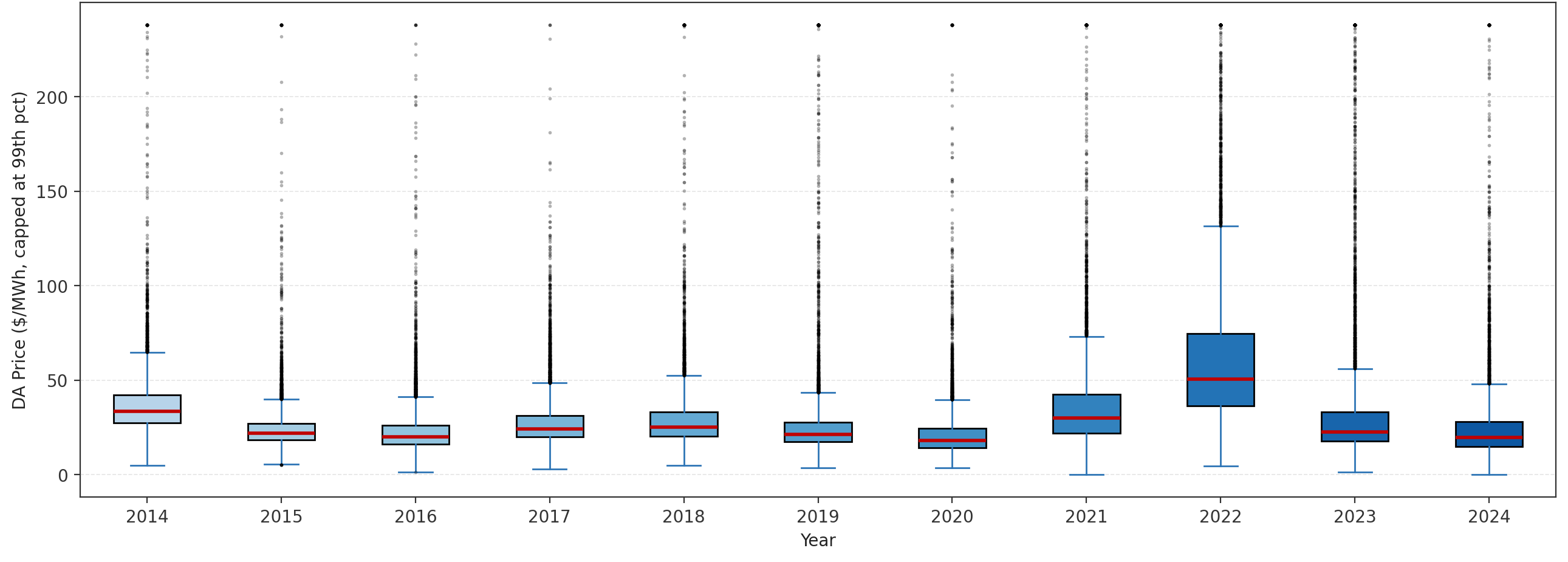}
        \vspace{-0.2cm}
        \caption{Yearly price distributions (2014-2024)}
        \label{fig:yearly_dist}
    \end{minipage}
\end{figure*}

% \begin{figure*}[!t]
% \centering
%     \centering
%     \includegraphics[width=0.75\textwidth]{fig03.png}
%     \vspace{-0.15cm} 
%     \caption{Average DA price heatmap: year versus month (\$/\text{MWh})}

% %\vspace{-0.15cm}   % İki şekil arasında boşluk

%     \centering
%     \includegraphics[width=0.75\textwidth]{fig04.png}
%     \vspace{-0.15cm} 
%     \caption{Average DA price heatmap: year versus hours of day (\$/\text{MWh}) }
%     \label{fig:XXXb}

% \end{figure*}

\begin{figure*}[!t]
    \centering
    \begin{minipage}{0.48\textwidth}
        \centering
        \includegraphics[width=\linewidth]{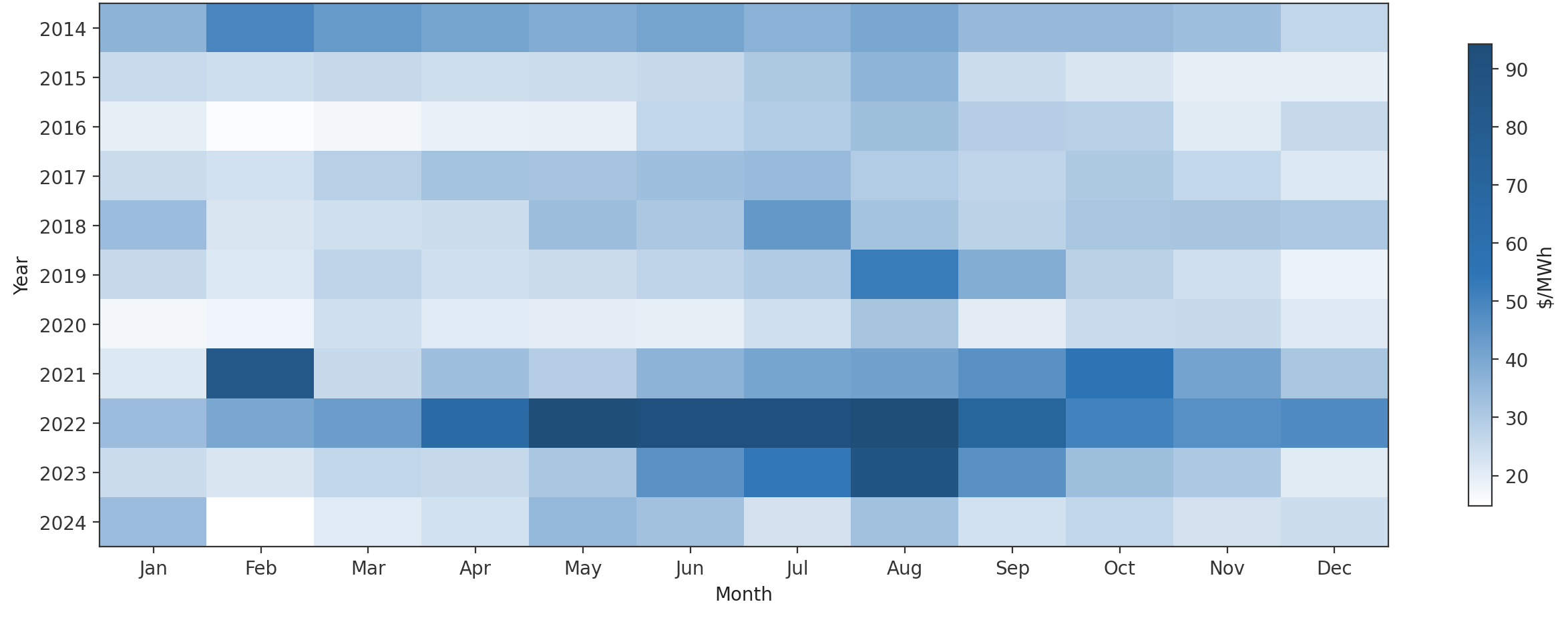}
        \vspace{-0.15cm}
        \caption{Average DA price heatmap: year versus month (\$/MWh)}
        \label{fig:heatmap_month}
    \end{minipage}
    \hfill
    \begin{minipage}{0.48\textwidth}
        \centering
        \includegraphics[width=\linewidth]{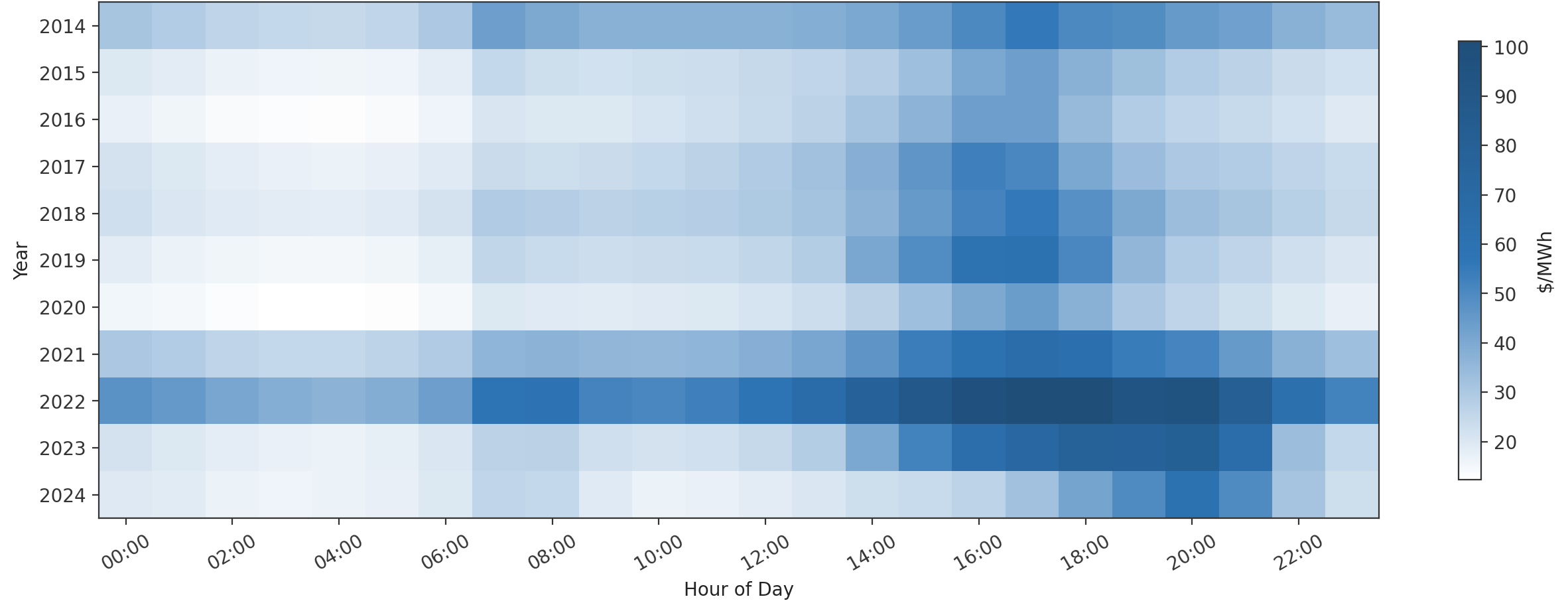}
        \vspace{-0.15cm}
        \caption{Average DA price heatmap: year versus hours of day (\$/MWh)}
        \label{fig:heatmap_hour}
    \end{minipage}
\end{figure*}

\section{Market Data Profile and Exploratory Observations}

\subsection{Market Trading Mechanisms}
The ERCOT electricity market is a system that is electrically isolated from the rest of the United States and lacks a capacity market. While structures such as Bilateral Agreements (BAs) defined by fixed prices agreed upon directly between two parties do exist, there are two key structures in which prices fluctuate according to market conditions. Trading in the Day-Ahead Market (DAM) is conducted on an hourly basis (with bidding closing at 10:00) and operates as a voluntary/financial mechanism. In the Real-Time Market (RTM), trading is possible in five-minute cycles and functions as a mandatory/physical mechanism. Naturally, in terms of trading volume, the DAM constitutes a significantly larger market, and its prices are generally more stable compared to those in the RTM. The long-term data to be used for this study was obtained from ERCOT DAM.

\subsection{Long-term Price Behavior}
An extended period spanning from 2014 to 2024 was evaluated within the scope of the HB-Houston location to observe general price fluctuations and boundary conditions in ERCOT-DAM prices. Price variations based on load, climatic data, and calendar tags were thoroughly analyzed during this period. Figure~\ref{fig:daily_max} shows the time series of daily maximum Day-Ahead (DA) prices spanning the period from 2014 to 2024, while the blue line represents the daily moving average. While maximum values generally remained stable until 2018, significant increases in price spikes were observed during the summer of 2019. Subsequently, the price surge that occurred in February 2021 (highlighted within the pink band) was recorded as the all-time high (reaching 8,000–9,000\$\text/{MWh}). As is evident from the graph, this value also had an impact on the moving average. Following this date, ERCOT implemented a series of measures; consequently, as of January 1, 2022, the price cap can no longer exceed \$5,000/\text{MWh}. During the years 2022 and 2023, a notable trend was the increased frequency and intensity of price spikes, particularly during the summer months. The year 2024, on the other hand, can be characterized as a year of normalization; during this period, both the frequency of spikes and the moving average values appear to have declined significantly.

Figure~\ref{fig:yearly_dist} shows the statistical price distribution on an annual basis. The central red line represents the median value, while the lower and upper edges of the box denote the 25th and 75th percentiles, respectively. To enhance the clarity of the graph, the upper limit of the price range has been capped at the 99th percentile. It is evident that, up until 2020, the concentration of prices fluctuated within a narrow range. However, starting from 2021 specifically during the years 2021 and 2022 prices are observed to have risen across a significantly wider band. When compared with Figure~\ref{fig:daily_max}, it can be argued that although the amplitude of the price spikes in 2022 was lower than in 2021 and 2023, the overall average remained higher with the increased frequency of these spikes.

Figure~\ref{fig:heatmap_month} presents the monthly distribution of average Day-Ahead (DA) prices in the form of a heatmap. February 2021 stands out distinctly from the other months. The impact felt due to the Uri shock which began that same year and the subsequent Russia-Ukraine war over the following two years can account for the darker areas visible in this graph. July and August appear to be the most volatile months in terms of pricing when viewed on a year over year basis. It is noted that, in 2024, the system may have become free of such price pressures. Among the potential factors contributing to this are stabilization of reforms, the expansion of solar and storage capacities aimed at ensuring supply-demand balance, and favorable climatic conditions. Figure~\ref{fig:heatmap_hour} presents the average hourly price behavior for the 2014–2024 period in the form of a heatmap. Until 2020, it is noted that there is no significant disparity among the hourly prices. However, in 2021 and 2022, there is a general increase in market prices across all hours. In 2023 and 2024, daytime hours become lighter in color, while evening hours remain dark. The impact of the increasing penetration of solar energy on prices in recent years can be interpreted as a duck curve.

%Apart from this, if price spikes are excluded, it can be said that generally speaking prices tend to rise in tandem with the annual increase in total load (r=0.211) over this 20 year period. While the standard deviation of annual prices exceeded \$850/MWh in 2021, the next closest value was calculated for 2023, standing slightly above \$200/MWh. As expected, weekend prices are generally lower than weekday prices; moreover, this differential widens further during the 06:00–09:00 and 13:00–19:00 time slots. However, between 00:00 and 04:00, weekend prices are albeit slightly higher than weekday prices. Specific patterns can also be observed, on both an hourly and monthly basis, regarding the frequency with which price spikes occur. When considering price spikes falling within the top 1\% percentile, it is observed that, on an hourly basis, they occur most frequently between 16:00 and 18:00. When this same analysis is performed on a monthly basis, the months with the highest frequency of occurrence in descending order are August, February, and July.

Beyond extreme (spike) events, day-ahead prices exhibit a moderate positive correlation with the annual growth in total system load ($r=0.211$) over the study period. Market volatility, however, has intensified significantly in recent years; while the standard deviation of annual prices peaked at over \$850/MWh in 2021 due to extreme weather events, the subsequent high of approximately \$200/MWh in 2023 underscores a persistent upward trend in price variance. Temporal analysis reveals distinct calendar-based patterns: weekday prices consistently exceed weekend levels, with this differential most pronounced during the morning (06:00--09:00) and afternoon (13:00--19:00) peaks. Conversely, a minor reversal is observed during the base-load hours of 00:00--04:00, where weekend prices marginally surpass those of weekdays. The distribution of price spikes—defined as the top 1\% of price observations—is similarly heterogeneous. Hourly analysis indicates a peak spike frequency between 16:00 and 18:00, aligning with the transition to evening net-load peaks. Seasonally, August, February, and July exhibit the highest incidence of spikes, reflecting the dual-peak demand nature of the ERCOT system driven by extreme cooling and heating requirements.

\section{Regime-Dependent
Explainable Analysis }

\subsection{Data Structure and Workflow}
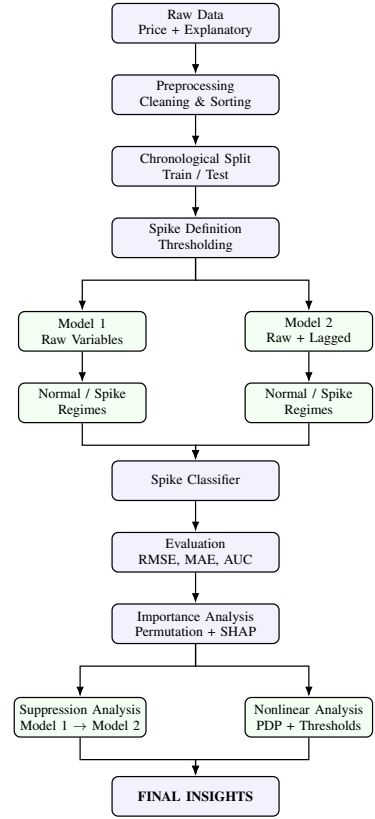
\begin{figure}[t]
\centering
\resizebox{0.27\textwidth}{!}{
\begin{tikzpicture}[
    node distance=7mm and 5mm, % Standard gap
    >=Latex,
    box/.style={
        rectangle,
        rounded corners,
        draw=black,
        thick,
        align=center,
        minimum width=3.8cm,
        minimum height=0.9cm,
        fill=blue!5,
        font=\small
    },
    smallbox/.style={
        rectangle,
        rounded corners,
        draw=black,
        thick,
        align=center,
        minimum width=2.9cm,
        minimum height=0.9cm,
        fill=green!5,
        font=\small
    },
    line/.style={draw, thick, -Latex}
]

% --- Central Top Pipeline ---
\node[box] (raw) {Raw Data\\Price + Explanatory};
\node[box, below=of raw] (prep) {Preprocessing\\Cleaning \& Sorting};
\node[box, below=of prep] (split) {Chronological Split\\Train / Test};
\node[box, below=of split] (spike) {Spike Definition\\Thresholding};

% --- First Parallel Block (Increased distance to 12mm to prevent overlap) ---
\node[smallbox, below left=12mm and -8mm of spike] (m1) {Model 1\\Raw Variables};
\node[smallbox, below right=12mm and -8mm of spike] (m2) {Model 2\\Raw + Lagged};

\node[smallbox, below=of m1] (reg1) {Normal / Spike\\Regimes};
\node[smallbox, below=of m2] (reg2) {Normal / Spike\\Regimes};

% --- Re-aligning to Center ---
% We use a coordinate to ensure the classifier is perfectly centered under 'spike'
\node[box, below=4.6cm of spike] (clf) {Spike Classifier};
\node[box, below=of clf] (eval) {Evaluation\\RMSE, MAE, AUC};
\node[box, below=of eval] (imp) {Importance Analysis\\Permutation + SHAP};

% --- Second Parallel Block ---
\node[smallbox, below left=12mm and -8mm of imp] (supp) {Suppression Analysis\\Model 1 $\rightarrow$ Model 2};
\node[smallbox, below right=12mm and -8mm of imp] (nonlin) {Nonlinear Analysis\\PDP + Thresholds};

% --- Final Box ---
\node[box, below=3cm of imp] (final) {\textbf{FINAL INSIGHTS}};

% --- Arrows with Straight-then-Turn Logic ---
\draw[line] (raw) -- (prep);
\draw[line] (prep) -- (split);
\draw[line] (split) -- (spike);

% Fork 1
\draw[line] (spike.south) -- ++(0,-0.5) -| (m1.north);
\draw[line] (spike.south) -- ++(0,-0.5) -| (m2.north);
\draw[line] (m1) -- (reg1);
\draw[line] (m2) -- (reg2);

% Join 1
\draw[line] (reg1.south) -- ++(0,-0.5) -| (clf.north);
\draw[line] (reg2.south) -- ++(0,-0.5) -| (clf.north);

\draw[line] (clf) -- (eval);
\draw[line] (eval) -- (imp);

% Fork 2
\draw[line] (imp.south) -- ++(0,-0.5) -| (supp.north);
\draw[line] (imp.south) -- ++(0,-0.5) -| (nonlin.north);

% Join 2
\draw[line] (supp.south) -- ++(0,-0.5) -| (final.north);
\draw[line] (nonlin.south) -- ++(0,-0.5) -| (final.north);

\end{tikzpicture}
}
\caption{Systematic workflow of the proposed methodology for ERCOT price analysis, isolating physical parameters from lag-based variables.}
\label{fig:flowchart}
\end{figure}

The data set used in this study was compiled from multiple publicly available sources, including GridStatus, the ERCOT Data Portal, and the National Renewable Energy Laboratory (NREL), integrating electricity prices, system load, weather observations, and calendar-related variables into a unified hourly time-series data set \cite{b27,b28,b29}. After synchronization, preprocessing, and quality control, the final dataset contained 95,303 hourly observations spanning January 2014 through December 2024.
To preserve the chronological nature of electricity markets and eliminate information leakage, the dataset was divided into a training period (2014–2023, 86,545 samples) and an independent test period (2024, 8,758 samples). Missing values were handled using median-value imputation before model training. The test dataset consisted of 8,646 normal observations and 112 spike observations, where spike events were defined according to the 98th percentile of the training price distribution.
Two different feature configurations were investigated. Model 1 utilized only physical and temporal explanatory variables (load, climate, and calendar features), whereas Model 2 additionally incorporated historical price-lag variables. The flowchart for executing the workflow is shown in Figure~\ref{fig:flowchart}. Table~\ref{tab:data_model_summary} summarizes the final dataset after preprocessing, the chronological train--test split, and the HGBR configuration used in the proposed framework. Reporting the exact sample sizes for the normal and spike subsets, along with the corresponding model settings, ensures full reproducibility and highlights the class imbalance inherent to price-spike prediction.

\vspace{-5pt}
\subsection{Mathematical and Statistical Framework}

\begin{table}[!t]
\caption{Summary of the dataset, chronological train--test partitioning, and HGBR model configuration.}
\label{tab:data_model_summary}
\centering
\renewcommand{\arraystretch}{1.05}
\begin{tabular}{ll}
\hline
\textbf{Property} & \textbf{Value} \\
\hline
Total observations & 95,303 \\
Training period & 2014--2023 \\
Training samples & 86,545 \\
Test period & 2024 \\
Test samples & 8,758 \\
Normal test observations & 8,646 \\
Spike test observations & 112 \\
Spike definition & 98th percentile of training prices \\
Machine learning model & HGBR \\
Learning rate & 0.05 \\
Maximum tree depth & 6 (Normal), 4 (Spike) \\
Maximum iterations & 300 (Normal), 250 (Spike) \\
Minimum samples per leaf & 30 (Normal), 10 (Spike) \\
Random state & 42 \\
\hline
\end{tabular}
\end{table}

The model is constructed using price, load, climate, and calendar variables as primary explanatory components reflecting the underlying physical and temporal structure of the electricity market, while lagged price variables are included only as a control to evaluate the statistical consistency and explanatory adequacy of the proposed framework. Here, the ERCOT day-ahead electricity price $P_t$ is the target variable, and $t=1,\dots, T$ denotes hourly observations. And the other explanatory variables are grouped into four different blocks: load, climate, Calendar, and price lag,
\begin{equation}
P_t := \text{DA\_price}_t.
   \end{equation}
\begin{itemize}
    \item Load block
    \begin{equation}
    \mathcal{X}^{(L)}_t = \{L_t\}
    \end{equation}

    \item Climate block
    \begin{equation}
    \mathcal{X}^{(C)}_t =
    \left\{
    \begin{array}{l}
    T_t,\ RH_t,\ WS_t,\\
    Pr_t,\ GHI_t
    \end{array}
    \right\}
    \end{equation}

    \item Calendar block
    \begin{equation}
    \mathcal{X}^{(K)}_t =
    \left\{
    \begin{array}{l}
    h_t,\ d^{(w)}_t,\ d^{(m)}_t,\ m_t,\\
    q_t,\ w_t,\ I^{\text{weekend}}_t,\\
    \sin_h(t),\ \cos_h(t),\ \sin_m(t),\ \cos_m(t)
    \end{array}
    \right\}
    \end{equation}

    \item Price lag block
    \begin{equation}
    \mathcal{X}^{(R)}_t =
    \left\{
    \begin{array}{l}
    P_{t-24},\ P_{t-48},\ P_{t-72},\\
    P_{t-168},\ P_{t-336},\\
    \bar{P}_{t-24},\ s_{t-24},\ \bar{P}_{t-168}
    \end{array}
    \right\}
    \end{equation}
\end{itemize}

where $L_t$ denotes the load, $T_t$ temperature, $RH_t$ relative humidity,
$WS_t$ wind speed, $Pr_t$ pressure, and $GHI_t$ global horizontal irradiance. Moreover, $h_t$ denotes hour, $d^{(w)}_t$ day of week, $d^{(m)}_t$ day of month, $m_t$ month, $q_t$ quarter, $w_t$ week of year, and $I^{\text{weekend}}_t$ the weekend indicator. Expressions containing $sin$ and $cos$ terms are used to represent the periodic nature of the corresponding time terms. Finally, ${P}_{t-24}$ and ${P}_{t-168}$ denote normal delays, $\bar{P}_{t-24}$ and $\bar{P}_{t-168}$ denote rolling means, and $s_{t-24}$ denotes rolling standard deviation over the previous 24 hours. Thus, the two main model input sets are:
\begin{equation} Model 1 \mapsto     
\mathcal{X}^{(1)}_t = \mathcal{X}^{(L)}_t \cup \mathcal{X}^{(C)}_t \cup \mathcal{X}^{(K)}_t,
\end{equation}
\begin{equation} Model2\mapsto
\mathcal{X}^{(2)}_t = \mathcal{X}^{(1)}_t \cup \mathcal{X}^{(R)}_t.
\end{equation}
The training period was defined as the years 2014–2023, and the test period as the year 2024. To identify extreme price conditions, a spike regime is defined using the $98^{\text{th}}$ percentile of the \emph{training} price distribution:
\begin{equation}
q_{0.98}^{\text{train}} = \text{Quantile}_{0.98}\big(\{P_t: t \in \mathcal{T}_{\text{train}}\}\big).
\end{equation}
The spike indicator is then defined as
\begin{equation}
S_t =
\begin{cases}
1, & \text{if } P_t > q_{0.98}^{\text{train}},\\[4pt]
0, & \text{otherwise}.
\end{cases}
\end{equation}
\subsection{Evaluation Metrics}
Two metrics are used to evaluate forecasting algorithms for time series in electricity markets; Mean Absolute Error (MAE), $\text{MAE} = \frac{1}{n}\sum_{t=1}^{n} |P_t - \widehat{P}_t|$, and Root Mean Squared Error (RMSE), $\text{RMSE} = \sqrt{\frac{1}{n}\sum_{t=1}^{n}(P_t - \widehat{P}_t)^2 }$.% usually stand out when evaluating forecasting algorithms for time series in electricity markets.

\subsection{Permutation Feature Importance}
PFI is a method used to measure in a model-agnostic manner how much a specific variable (or feature) truly contributes to a model
\begin{equation}
I_j^{\text{perm}} = \mathcal{L}(f,\pi_j(\mathcal{D})) - \mathcal{L}(f,\mathcal{D}).
\end{equation}
Where $f$ is a trained model and  $x_j$ is a feature. Permutation importance $I_j^{\text{perm}}$ is computed by randomly permuting the values of $x_j$ in the evaluation set and measuring the loss increase. $\mathcal{L}(f,\mathcal{D})$ denotes the evaluation loss on dataset $\mathcal{D}$ and $\pi_j(\mathcal{D})$ denotes the dataset where feature $x_j$ is randomly permuted. A larger value of $I_j^{\text{perm}}$ indicates that feature $x_j$ is more important to model performance. 

For comparison purposes, feature importance scores are normalized according to $w_j = \frac{|I_j|}{\sum_{k=1}^{p}|I_k|}$ and $0 \leq w_j \leq 1, \qquad \sum_{j=1}^{p} w_j = 1$. Then, the calculated PFI is used to measure how much of the explanatory power of other factors is absorbed by the price lag: 
\begin{equation}
\text{SR}_j
=\frac{w_j^{(1)} - w_j^{(2)}}{w_j^{(1)}}.
\end{equation}
Where $w_j^{(1)}$ denotes the normalized importance of feature $j$ in Model 1, and let $w_j^{(2)}$ denotes the normalized importance of the same feature in Model 2. Accordingly, $\text{SR}_j > 0$: the feature is weakened after adding lag. $\text{SR}_j \approx 1$: the feature is almost fully absorbed by lag. $\text{SR}_j \approx 0$: the feature is mostly preserved. $\text{SR}_j < 0$: the feature becomes relatively more important after adding lag.

%\\ (T1:-7.4,-2.2; T2:-2.2,22.75; T3:22.75,27.30; T4:27.30,32.1), (H1:26.62,58.63; H2:58.63,74.23; H3:74.23,90.03; H4:90.03,100)
If the features are treated as a group $\mathcal{J}_G$, group (as a block) importance and block-level suppression ratio $SR$ are defined as
\begin{equation}
W_G = \sum_{j \in \mathcal{J}_G} w_j.
\end{equation}
\begin{equation}
\text{SR}_G
= \frac{W_G^{(1)} - W_G^{(2)}}{W_G^{(1)}}.
\end{equation}

\section{Results and Discussion }
% \begin{figure*}[h!]
%     \centering
%     % Row 1
%     \begin{subfigure}{0.47\textwidth}
%         \centering
%         \includegraphics[width=\linewidth]{a1.PNG}
%         \caption{Normal regime - Model 1 (No Lags)}
%         \label{fig:norm_phys}
%     \end{subfigure} \hfill
%     \begin{subfigure}{0.47\textwidth}
%         \centering
%         \includegraphics[width=\linewidth]{b1.PNG}
%         \caption{Normal regime - Model 2 (With Lags)}
%         \label{fig:norm_lags}
%     \end{subfigure}
    
%     \vspace{6pt} % Slight gap for better readability between rows
    
%     % Row 2
%     \begin{subfigure}{0.47\textwidth}
%         \centering
%         \includegraphics[width=\linewidth]{c1.PNG}
%         \caption{Spike regime - Model 1 (No Lags)}
%         \label{fig:spike_phys}
%     \end{subfigure} \hfill
%     \begin{subfigure}{0.47\textwidth}
%         \centering
%         \includegraphics[width=\linewidth]{d1.PNG}
%         \caption{Spike regime - Model 2 (With Lags)}
%         \label{fig:spike_lags}
%     \end{subfigure}

%     \caption{Comparison of Permutation Feature Importance (PFI) across regimes. (a) and (b) illustrate the masking effect of price lags under normal conditions, while (c) and (d) highlight the breakdown of lag-based importance during price spikes.}
%     \label{fig:importance_panel}
% \end{figure*}

\begin{figure*}[htbp]
    \centering
    % Row 1
    \subfloat[Normal regime - Model 1 (No Lags)\label{fig:norm_phys}]{%
        \includegraphics[width=0.47\textwidth]{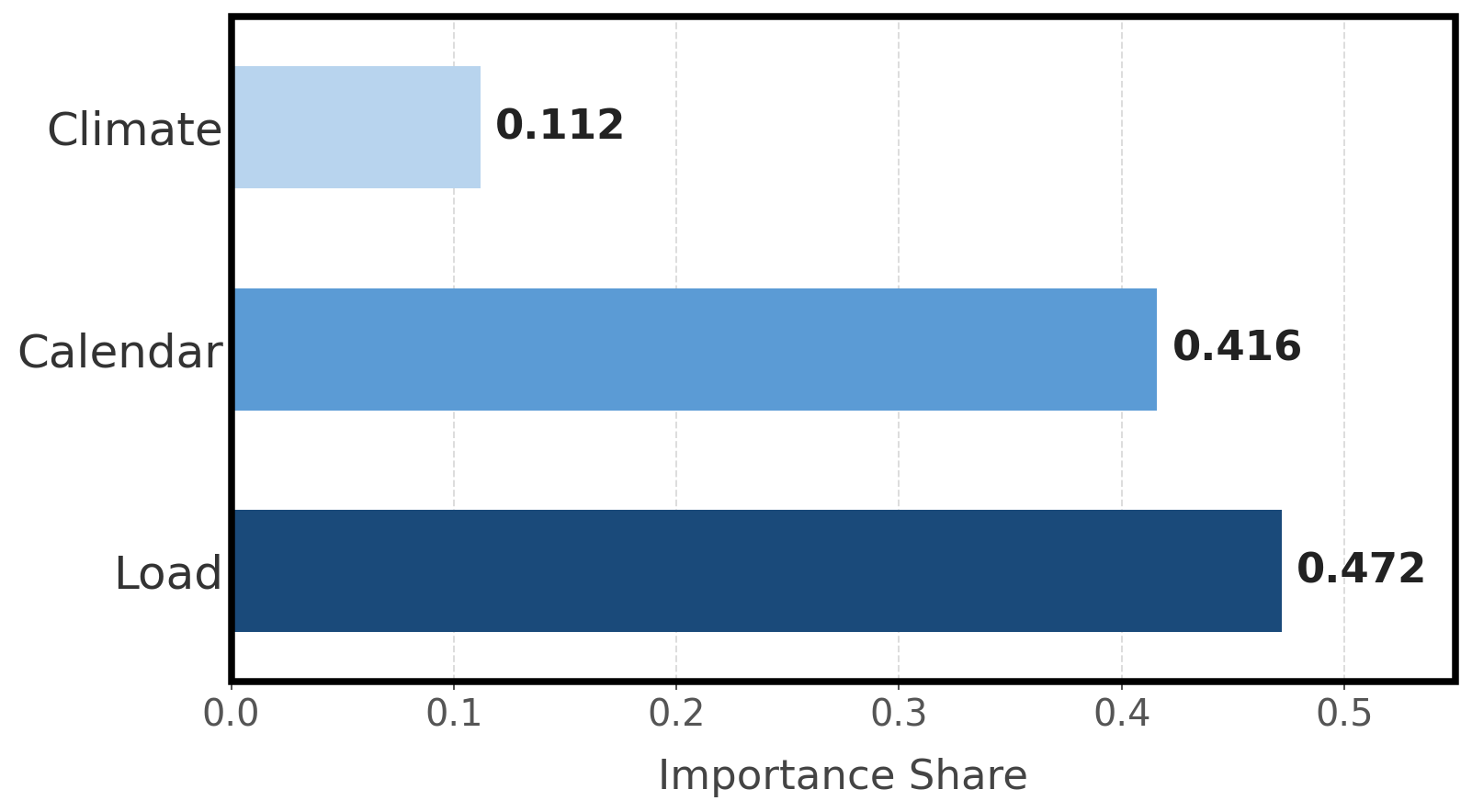}%
    }%
    \hfill
    \subfloat[Normal regime - Model 2 (With Lags)\label{fig:norm_lags}]{%
        \includegraphics[width=0.47\textwidth]{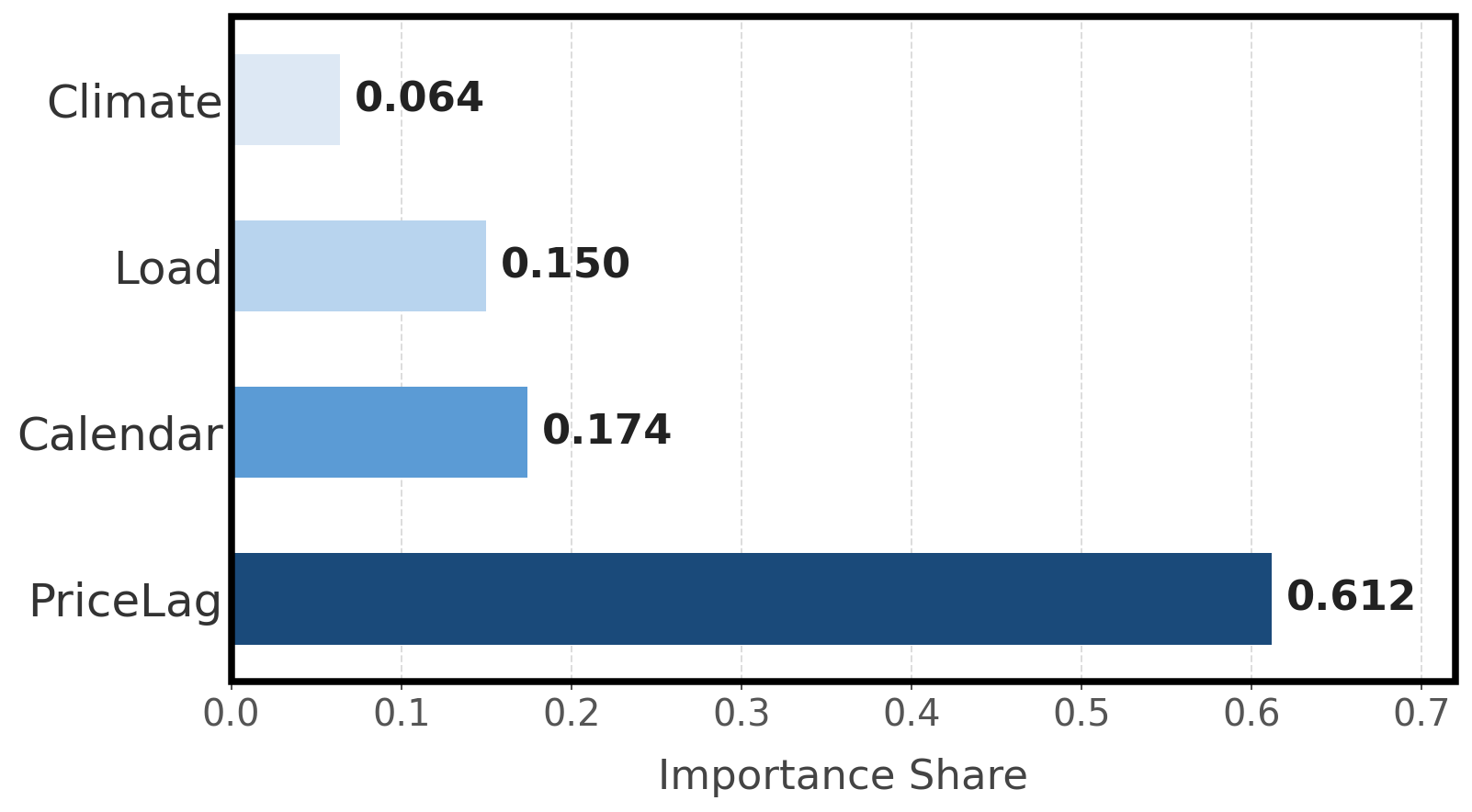}%
    }
    
    \vspace{6pt} % Gap between rows
    
    % Row 2
    \subfloat[Spike regime - Model 1 (No Lags)\label{fig:spike_phys}]{%
        \includegraphics[width=0.47\textwidth]{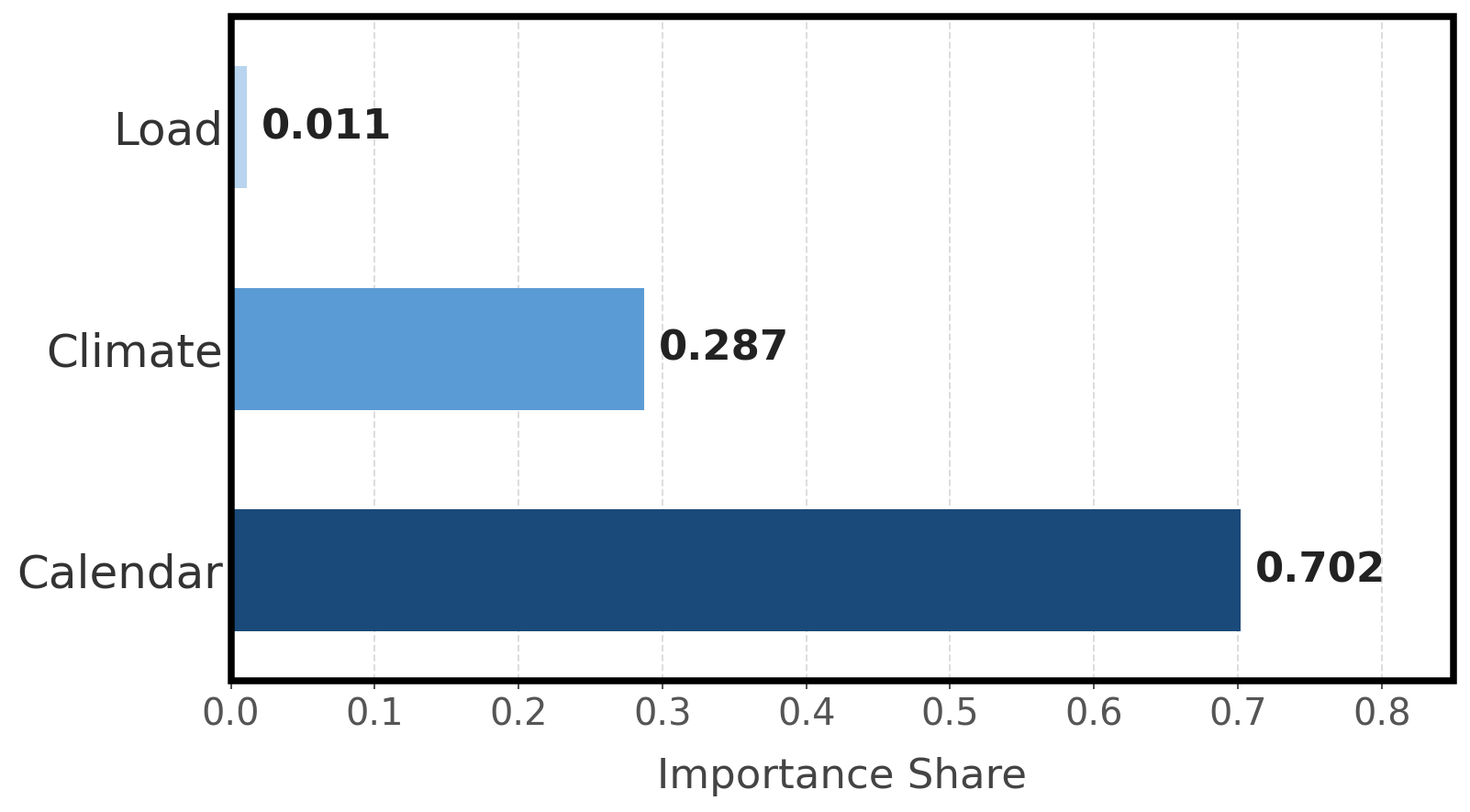}%
    }%
    \hfill
    \subfloat[Spike regime - Model 2 (With Lags)\label{fig:spike_lags}]{%
        \includegraphics[width=0.47\textwidth]{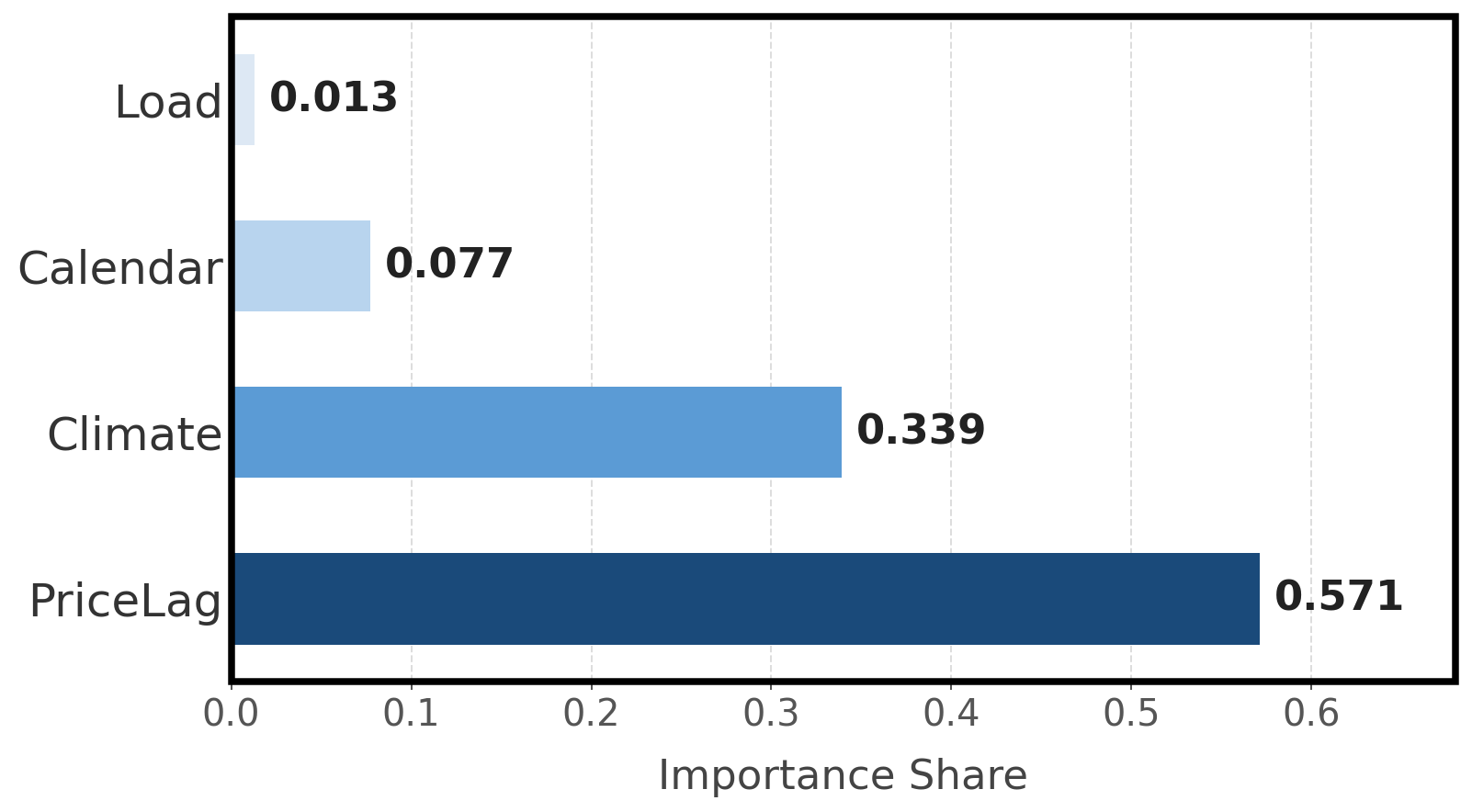}%
    }

    \caption{Comparison of Permutation Feature Importance (PFI) across regimes. (a) and (b) illustrate the masking effect of price lags under normal conditions, while (c) and (d) highlight the breakdown of lag-based importance during price spikes.}
    \label{fig:importance_panel}
\end{figure*}
Figure~\ref{fig:importance_panel} generally illustrates regime-dependent price formation. Accordingly, under normal conditions, price behavior is shaped by load and calendar effects, with climatic effects also playing a role to a certain extent, as depicted in Figure 6(a). When a lag term is introduced, the "memory" associated with prices becomes the dominant factor. In fact, the price lag term does not merely represent the past price; rather, it encapsulates a compressed summary of various features—including load, calendar, and climatic characteristics—alongside numerous other attributes. This phenomenon is evident in Figure~\ref{fig:importance_panel}(b), where the relative importance of the Load and Calendar variables decreases significantly. Based on data obtained from the test set, the error metric for Model 1 stood at 14.1, whereas for Model 2, it dropped to 6.39.

\begin{table}[b!]
\centering
\caption{Suppression ratios }
\label{tab:suppression}
\begin{tabular}{lcc}
\hline
\textbf{Block} & \textbf{Normal Regime} & \textbf{Spike Regime} \\
               & \textit{\scriptsize (Model 1 $\rightarrow$ Model 2)} & \textit{\scriptsize (Model 1 $\rightarrow$ Model 2)} \\
\hline
Load     & 0.681  & -0.207 \\
Climate  & 0.425  & -0.182 \\
Calendar & 0.582  & 0.891 \\
\hline
\end{tabular}
\end{table}

In the spike regime, Model 1 identifies temporal identifiers (calendar variables) as the primary drivers of price formation, with climatic factors showing secondary significance; notably, system load appears to offer negligible marginal importance. However, the introduction of the price-lag variable in Model 2 induces a significant suppression effect: the influence of the calendar is largely subsumed, while the relative importance of climate and load increases, as illustrated in Figures~\ref{fig:importance_panel}(c) and \ref{fig:importance_panel}(d). Table II presents the suppression rates associated with the transition from Model 1 to Model 2 for both the normal and spike price regimes. Specifically, within the spike regime, the inclusion of price lags did not positively impact performance; indeed, an increase in both MAE and RMSE values was observed in Model 2 compared to Model 1. This outcome clearly demonstrates that the addition of lag variables is insufficient on its own to accurately determine price levels of spikes.

Regime-dependent diagnostic performances of the proposed models are presented comparatively in Table~\ref{tab:diagnostic}. Since the primary focus of this study is not to improve or optimize prediction accuracy, the RMSE and MAE values presented in Table~\ref{tab:diagnostic} should not be interpreted as competitive forecasting benchmarks. An important observation emerging from Table~\ref{tab:suppression} and Table~\ref{tab:diagnostic} is that predictive performance and model interpretability do not necessarily evolve in parallel. Under spike conditions, the relatively small variation in conventional error metrics could easily be interpreted as evidence that lag variables have little influence on model behavior. However, Table~II demonstrates that the same lag variables considerably reduce the apparent contribution of load and climate features. This discrepancy suggests that conventional forecasting metrics alone are insufficient to evaluate the explanatory behavior of electricity price forecasting models.

\begin{figure}[!t]
\centering
\includegraphics[width=0.9\columnwidth]{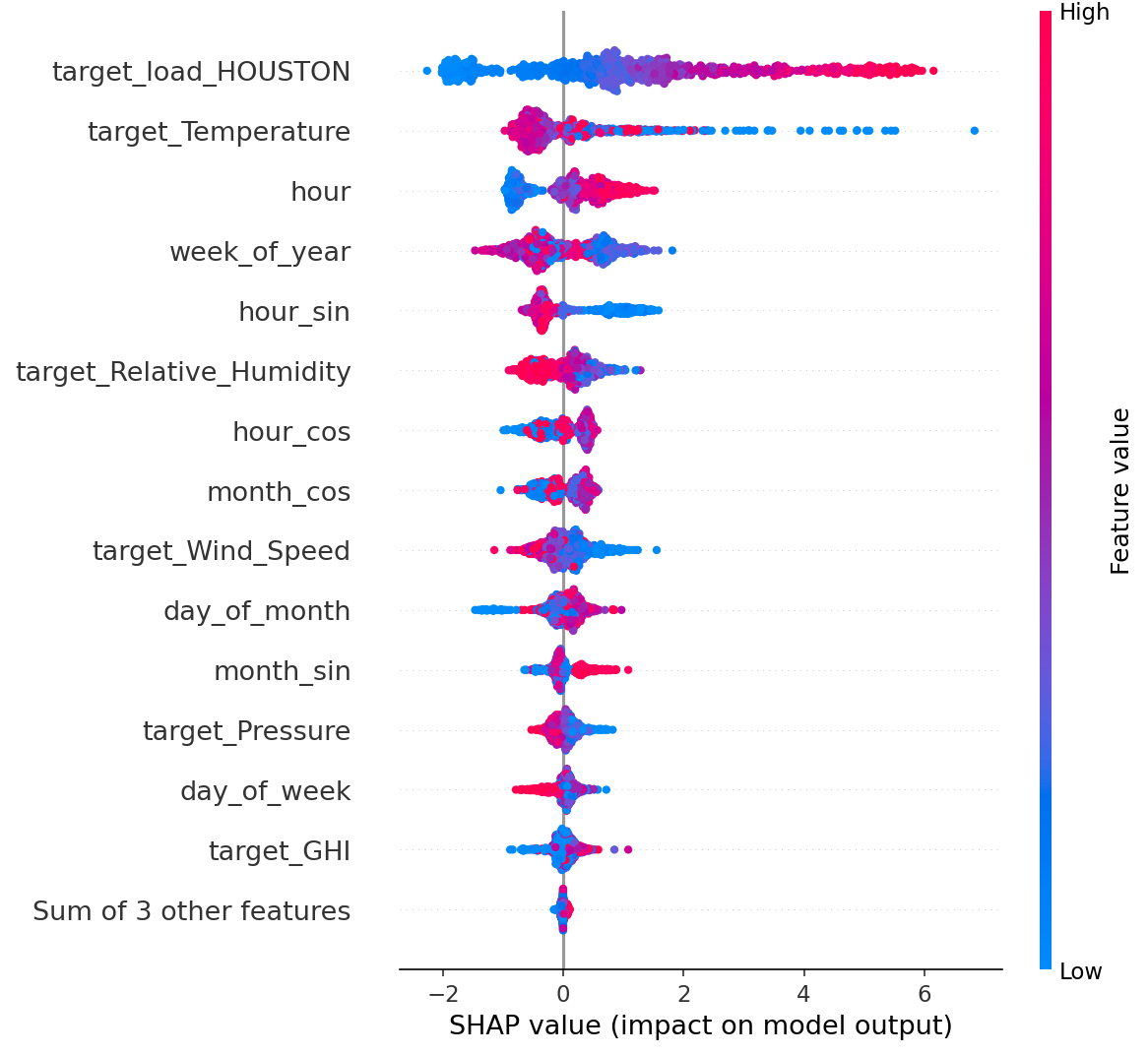}
\caption{SHAP beeswarm plot for spike occurrence classifier.}
\label{fig:shap_beeswarm}
\vspace{-15pt}
\end{figure}

With respect to the 2024 test data, the model is capable of detecting the occurrence of a price spike with a probability of 91.5\%. However, the actual magnitude of such a spike price requires evaluation using a distinct set of parameters. The SHAP analysis presented in Figure~\ref{fig:shap_beeswarm} naturally indicates that system load is the factor most likely to trigger significant price spikes. However, it also demonstrates that this likelihood is closely linked to temporal variables and that temperature can play a significant role during periods of both extreme cold and extreme heat.

\begin{table*}[!h]
\caption{Regime-dependent performance of the proposed models.}
\label{tab:diagnostic}
\centering
\renewcommand{\arraystretch}{1.2}
\setlength{\tabcolsep}{7pt}

\begin{tabular}{l p{7.2cm} p{7.2cm}}
\hline
\textbf{Metric} &
\centering\textbf{Normal Regime (Model 1 $\rightarrow$ Model 2)} &
\centering\textbf{Spike Regime (Model 1 $\rightarrow$ Model 2)}
\tabularnewline
\hline

\textbf{RMSE} &
\centering 18.07 $\rightarrow$ 10.31 (-42.9\%) &
\centering 443.06 $\rightarrow$ 451.62 (+1.9\%)
\tabularnewline

\textbf{MAE} &
\centering 14.10 $\rightarrow$ 6.39 (-54.7\%) &
\centering 299.31 $\rightarrow$ 316.79 (+5.8\%)
\tabularnewline

\textbf{Finding 1} &
Significant reduction in RMSE and MAE. &
Only marginal increase in RMSE and MAE.
\tabularnewline

\textbf{Finding 2} &
Strong temporal persistence is captured by lag variables. &
Historical price lags provide limited additional information.
\tabularnewline

\hline
\end{tabular}
\end{table*}

%A key finding was the identification of a distinct non-linear relationship between temperature and humidity during sudden price surges, as shown in Figure~\ref{fig:matrix1}. The highest prices occurred within the moderate temperature range ($T3$) and under high humidity conditions ($H4$). Significantly lower prices were observed under similar humidity conditions, but at higher temperature levels. The temperature-price relationship is not monotonic, and the humidity factor exhibits a multiplier effect that only drives up prices within specific temperature ranges. The interaction of these two important climatic parameters could be one of the elements that can be used to explain spike behavior. In Figure~\ref{fig:tempthresholds}, these curves serve to corroborate the preceding observation. Accordingly, it is clearly evident that both normal and spike price values tend to decline once temperature increases surpass a specific range. For spike prices, this critical range appears to lie approximately between 24°C and 27°C. Consequently, this supports that price fluctuations are driven by a threshold-based mechanism rather than by monotonic temperature effects.

Following the SHAP-based analysis, temperature and relative humidity emerged as two of the most influential meteorological variables affecting electricity price formation after system load. To further examine how these variables jointly influence market behavior, their interaction was analyzed using the complete 2014-2024 dataset while excluding the extraordinary 2021 Texas Freeze event. This extended analysis provides a broader characterization of the meteorological conditions associated with price formation than would be obtained from a single evaluation year alone. The temperature and relative humidity values have initially been divided into four equal-frequency groups (quartiles). The sample counts shown in Figure 8 represent the number of observations in each temperature-humidity group following application of the spike price regime filter. The resulting sample counts are sufficiently large for most temperature–humidity groups and are shown directly in Figure 8.  As shown in Figure 8, the effect of temperature on electricity prices varies depending on different humidity levels; this indicates a nonlinear relationship between these two meteorological variables.

Using the same meteorological data set, the long-term temperature response has been analyzed to determine whether the effect of temperature exhibits a different transition behavior or follows a monotonic trend. The temperature-price relationship cannot be sufficiently explained by a single pattern, as Figure 9 shows that temperature affects electricity prices differentially under normal and spike price conditions. Under the normal price regime, average prices follow a relatively stable trajectory at moderate temperatures, whereas they gradually rise to approximately \$46.4/MWh in the highest temperature range ($\approx$39.6$^{\circ}$C). In contrast, the spike price regime exhibits substantially greater price variability, with average prices increasing rapidly beyond approximately 30°C and reaching nearly \$715/MWh around 35.7°C. This pattern indicates that spike price occurrences react in a different way to temperature compared to normal market prices, which is in line with rising cooling demand, narrowing operating margins, and increased supply–demand stress during hot weather.

\begin{figure}[!t]
\centering
\includegraphics[width=1\columnwidth]{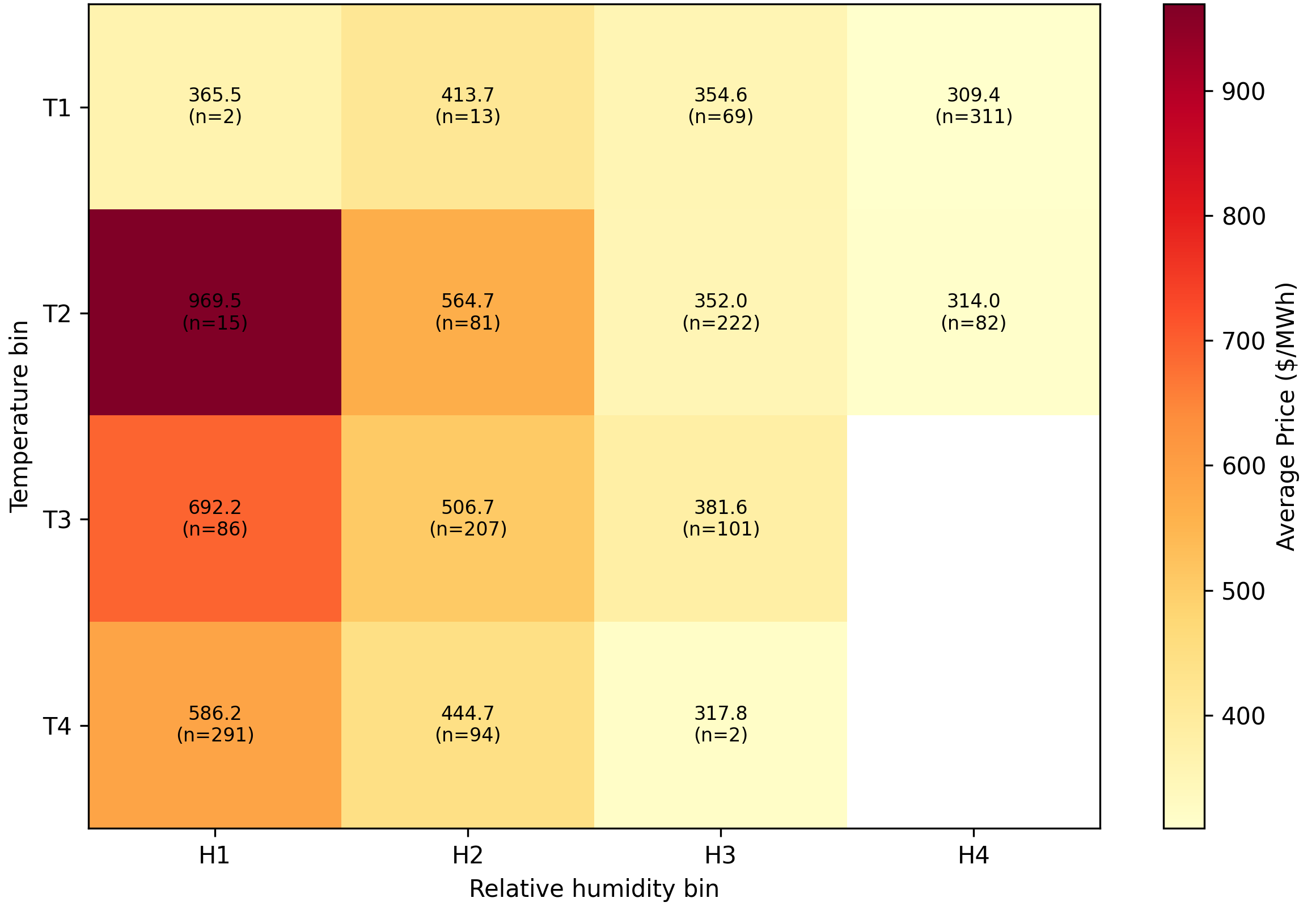}
\caption{Long term  spike regime between 2014 and 2024:\\ temperature and humidity interaction, except for 2021.}
\label{fig:matrix1}
\vspace{-10pt}
\end{figure}

\begin{figure}[!t]
\centering
\includegraphics[width=0.95\columnwidth]{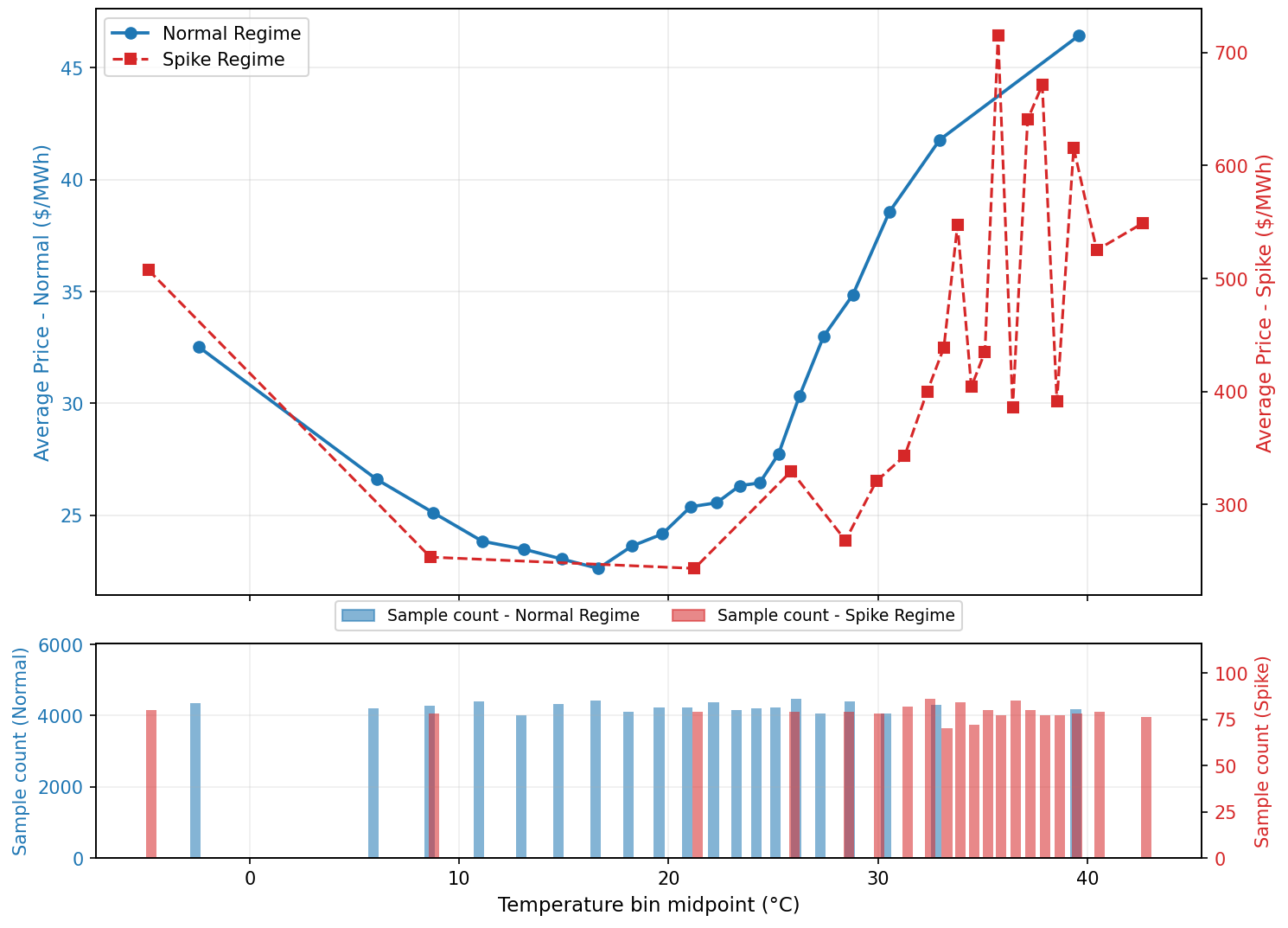}
\caption{Long term  spike regime between 2014 and 2024:\\ Temperature thresholds vs Avg. prices, except for 2021.}
\label{fig:tempthresholds}
\vspace{-10pt}
\end{figure}

\section{Conclusion}
This study analyzes, based on price regimes, how and to what extent lagged prices mask the true drivers of electricity price dynamics. The results indicate that, under normal conditions, lagged prices capture a significant portion of the temporal and demand-related structure; however, because they obscure key physical mechanisms, they fail to explain critical price spikes. Once this masking effect is eliminated, a regime-dependent structure becomes apparent. The pattern emerging from long-term meteorological analyses shows that electricity prices exhibit distinctly different behaviors depending on temperature under normal and spike price increase regimes. In addition, the interaction between temperature and relative humidity exhibits a significantly nonlinear nature. Naturally, these findings are a product of variables specific to the local region where the study was conducted and, therefore, cannot be generalized as a universal market rule. However, it's possible that similar patterns could emerge in other markets and regions using the same research methodology.

\section*{Acknowledgment}
This study was supported by The Scientific and Technological Research Council of Türkiye (TÜBİTAK-2219) International Postdoctoral Research Fellowship No. 1059B192402447. Various AI tools were used to improve the narrative language.

% Please number citations consecutively within brackets \cite{b1}. The 
% sentence punctuation follows the bracket \cite{b2}. Refer simply to the reference 
% number, as in \cite{b3}---do not use ``Ref. \cite{b3}'' or ``reference \cite{b3}'' except at 
% the beginning of a sentence: ``Reference \cite{b3} was the first $\ldots$''

% Number footnotes separately in superscripts. Place the actual footnote at 
% the bottom of the column in which it was cited. Do not put footnotes in the 
% abstract or reference list. Use letters for table footnotes.

% Unless there are six authors or more give all authors' names; do not use 
% ``et al.''. Papers that have not been published, even if they have been 
% submitted for publication, should be cited as ``unpublished'' \cite{b4}. Papers 
% that have been accepted for publication should be cited as ``in press'' \cite{b5}. 
% Capitalize only the first word in a paper title, except for proper nouns and 
% element symbols.

% For papers published in translation journals, please give the English 
% citation first, followed by the original foreign-language citation \cite{b6}.

\end{document}